\documentclass[aip,pop,amsmath,amssymb,reprint]{revtex4-1}
\usepackage{graphicx}

\begin{document}

\title[Scanning drift tube for mapping electron swarms]{A scanning drift tube apparatus for spatio-temporal mapping of electron swarms}

\author{I. Korolov} 
\affiliation{Institute for Solid State Physics and Optics, Wigner Research Centre for Physics, Hungarian Academy of Sciences, P.O.B. 49, H-1525 Budapest, Hungary}

\author{M. Vass}
\affiliation{Institute for Solid State Physics and Optics, Wigner Research Centre for Physics, Hungarian Academy of Sciences, P.O.B. 49, H-1525 Budapest, Hungary}

\author{N. Kh. Bastykova}
\affiliation{Institute of Experimental and Theoretical Physics, al-Farabi Kazakh National University, 96a Tole Bi, Almaty 050012, Kazakhstan}

\author{Z. Donk\'o}
\email{donko.zoltan@wigner.mta.hu}
\affiliation{Institute for Solid State Physics and Optics, Wigner Research Centre for Physics, Hungarian Academy of Sciences, P.O.B. 49, H-1525 Budapest, Hungary}

\date{\today}

\begin{abstract}
A ``scanning'' drift tube apparatus, capable of mapping of the spatio-temporal evolution of electron swarms, developing between two plane electrodes under the effect of a homogeneous electric field, is presented. The electron swarms are initiated by photoelectron pulses and the temporal distributions of the electron flux are recorded while the electrode gap length (at a fixed electric field strength) is varied. Operation of the system is tested and verified with argon gas, the measured data are used for the evaluation of the electron bulk drift velocity. The experimental results for the space-time maps of the electron swarms -- presented here for the first time -- also allow clear observation of deviations from hydrodynamic transport. The swarm maps are also reproduced by particle simulations.
\end{abstract}  

\pacs{52.65.-y, 52.25.Fi, 52.25.-b}

\keywords{electron swarms, particle kinetics, numerical simulation}
  
\maketitle

\section{Introduction}

Transport coefficients of charged particles in a background gas, like the drift velocity, mean energy, and diffusion coefficients, are fundamentally important quantities in {\it swarm physics} and serve as basic input data in {\it gas discharge modeling}. Additionally, they allow checking and adjusting the cross section sets of collision processes, as the transport coefficients can accurately be computed from the cross section data.\cite{ZP} 

Measurements of the transport coefficients can be realized in experiments, where a low-density ensemble of charged particles moves under the influence of an (ideally homogeneous) electric field. Such ``swarm experiments'', carried out in {\it drift tubes}, have attracted a significant attention throughout the past decades. Most of the work has concentrated on measurements of electron and ion transport coefficients \cite{D1,D2,D3}, however, more recently positron transport turned into the focus of interest as well.\cite{positron1,positron2,positron3} 

Current developments in the field are motivated by several factors: (i) the increasing purity of the experimental apparatuses and the improvements of the data acquisition methods make it possible to refine previously available experimental data; (ii) the need for information about particle transport in gases that became important more recently due to their applications in plasma processing (e.g. CF$_4$, \cite{cf4} CF$_3$I,\cite{cf3i_1} tetraethoxysilane \cite{TEOS}), due to their relevance to global warming (e.g. SF$_6$ \cite{sf6_2,cf3i_2}) and to biomedicine (He+H$_2$O mixtures\cite{J_h2o}), etc.; (iii) the fundamental interest in transport phenomena in specific gas mixtures where particular effects, like negative differential conductivity, take place.\cite{P_ndc,U_ndc,Nikolay} 

Drift tubes usually operate at a fixed electrode gap length, or at a few discrete values of the gap length. Thus, the properties of the electron swarms are sampled at one, or at most, a few positions only, and different approximations have to be adopted to derive the transport parameters from the measured signals. Drift tubes, could, on the other hand, provide information about the {\it spatial and temporal evolution of the swarms}, provided that data are recorded at a large number of electrode separations. The purpose of this paper is to report the development of such an apparatus and to demonstrate its unique capability of providing ``complete'' maps of the electrons swarms. 

\begin{figure*}[t!]
\includegraphics[width=0.9\textwidth]{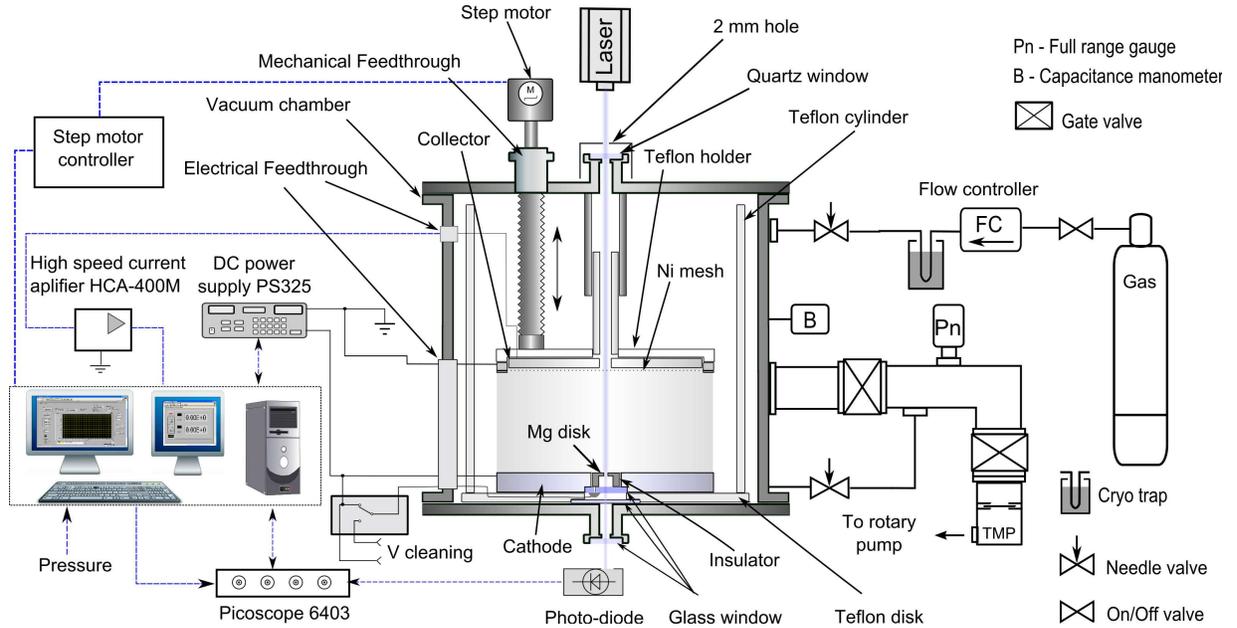}
\caption{\label{fig:expsetup}
(color online) Schematic of the scanning drift tube apparatus, the vacuum and control systems.}
\end{figure*}

For an accurate determination of the transport coefficients it is, in general, essential that drift tubes are operated in the {\it hydrodynamic regime} where transport coefficients depend only on the reduced electric field (electric field to gas density ratio), $E/N$, otherwise the results will be influenced by the actual conditions of the measurements and the dimensions of the cell.\cite{ZP} With a scanning system, the analysis of the measured maps allows one to directly distinguish between conditions of hydrodynamic and non-hydrodynamic transport. Due to this fact, (i) derivation of the transport parameters can be constrained to conditions where transport is in the hydrodynamic regime, and (ii) the equilibration of the swarm -- being an interesting phenomenon itself -- can directly be ``seen'' and studied.  
  
We demonstrate the operation of the system by recording maps of electron swarms in argon gas, for a wide range of $E/N$ values. Our system -- as will be explained later -- senses the conduction current of the electron swarm, and the data analysis follows clear steps to the determination of the {\it bulk drift velocity} $W$ (the velocity of the center-of-mass of the electron cloud). The measured values of $W$ are found to be in good agreement with previous experimental results. To complement the measurements of swarm maps, particle simulations (based on the Monte Carlo method) are also conducted and the results are compared with our experimental observations.  

The paper is organized as follows. In section II we describe the details and the operation principle of our experimental apparatus. A brief description of the computational approach is given in section III. The experimental and simulation results are presented and compared in section IV, while a summary of the work is given in section V.

\section{Experimental setup}

\subsection{Construction details and operation method}

The scheme of our experimental setup is shown in Figure~\ref{fig:expsetup}. The drift cell is situated within a stainless steel vacuum chamber and operates with photoelectron pulses, similarly to other more recent experimental setups.\cite{cf3i_1,zurich}  
The electrodes are made of stainless steel. The negatively biased cathode is situated at the bottom of the cell. To produce photoelectrons we use a frequency-quadrupled diode-pumped MPL-F-266 nm laser with singe-pulse energy of 1.7 $\mu$J ($\lambda=266$ nm) at 3 kHz repetition rate. The top flange of the chamber is equipped with a quartz window and a 2-mm diameter diaphragm that allows the laser light to enter the vacuum chamber. The upper (grounded) electrode has a hole with a diameter of 5 mm to let the laser beam pass trough the electrode gap and to illuminate a magnesium disk with 5 mm diameter and thickness of 4 mm that is embedded inside the cathode. Mg is chosen due to its relatively high quantum efficiency,\cite{Wang1995} $\eta_{\rm Mg} \approx 5 \times 10^{-4}$. The disk is separated from the cathode by a Teflon insulator and is connected to an electrical feedthrough with a Kapton wire. This configuration allows us to clean the surface of the Mg disk without opening the chamber by applying a DC voltage and creating a discharge between the disk and the cathode. This cleaning procedure is routinely applied prior to the measurements. During the course of the measurements the Mg disk and the cathode are connected. The Mg disk has a hole of 0.3 mm in diameter to allow part of the laser light to be detected by a photo-diode (with a fast response time of 0.39 ns) placed underneath the cell, outside the vacuum chamber. The signal of this photodiode is used as the trigger for the measurements.
A Teflon disk placed under the cathode and a Teflon cylinder that surrounds the drift cell prevent gas breakdown between the drift tube and the grounded chamber wall during operation when potentials up to several hundred Volts are applied to the cathode.      

Preceding the measurements the chamber is pumped down to a base pressure of $10^{-5}$ Pa (for several days). In the present experiment, we use 6.0 purity Ar (with impurities H$_2$O$\leq$0.5 ppm, O$_2 \leq 0.5$ ppm, C$_n$H$_m\leq 0.1$ ppm, CO$_2$ $\leq 0.1$ ppm, CO $\leq 0.1$ ppm, N$_2 \leq 0.5$ ppm, H$_2 \leq 0.5$ ppm), at a slow (2-5 sccm) flow, regulated by a flow controller. Prior to entering the chamber, to remove any remaining impurities, the gas is fed through a cryogenic (liquid nitrogen) trap. The gas pressure is measured by a Pfeiffer CMR 362 capacitive gauge with 0-1000 Pa range. The experiments have been carried out at a temperature of $T$ = 293 $\pm$ 2 K.

The collector, at the top of the cell, has a diameter of 105 mm and is fixed with screws to a Teflon holder. An electrically grounded nickel mesh (grid) with $T$ = 88\% transmission and 45 lines/inch density (MN17 type, manufactured by Precision Eforming LLC) is situated at a distance 1 mm in front of the (grounded) collector electrode. The distance between the cathode and the grid can be set within the range of $L$ = 13.6 -- 63.6 mm, via a vacuum feedthrough equipped with a micrometer screw attached to a stepping motor. The signal from the collector -- generated by the flux of electrons entering the grid-collector gap (see later) -- is fed to a high speed current amplifier (type Femto HCA-400M) and is measured, synchronized with the photo-diode signal, by a digital oscilloscope (type Picoscope 6403B) with 0.8 ns time resolution. This time dependent signal (``electron arrival spectrum'') following each laser pulse is stored and is subsequently averaged for 3000-5000 pulses (depending on the signal intensity), at each position. As the detected signal is usually in the nA range, proper shielding of the electrical connections is required. 

The experiment is fully controlled by a computer using LabView software. The electrode gap is scanned over the whole range with typically 0.2 -- 1 mm step size in sequential measurements, while an accelerating voltage (cathode potential) is set by a PS-325 (Stanford Research Systems) power supply connected to the computer via a GPIB interface. During the mapping sequence, the voltage is set to ensure that $E/N$ remains fixed for the different gap sizes. Scanning the electrode separation and recording the arrival spectrum of the electrons at the collector allows us to obtain complete information about the {\it spatial and temporal evolution of the electron swarms}. Recording of swarm maps, with an acceptable signal to noise ratio, takes $\approx$25--100 minutes. To ensure accurate results (i) the time-dependence of the intensity of the laser is recorded during the whole course of the measurements and (ii) following the complete scan of electrode gaps an additional measurement is carried out at the starting position. Drift velocity values measured on different days did not differ more than 2\%, confirming the stability of the apparatus. 

\subsection{Operation principle and data interpretation}

In the hydrodynamic regime the space- and time-dependence of the electron density of a swarm generated at $t=0$ and $z=0$ is given as (see, e.g. Ref.\citenum{R}):
\begin{equation}
n(z,t) = \frac{n_0}{(4 \pi D_L t)^{1/2}} \exp \biggl[ \nu_{\rm i} t - \frac{(z - Wt)^2}{4 D_L t} \biggr],
\label{eq:n}
\end{equation}
where $n_0$ is the initial electron density, $D_L$ is the longitudinal diffusion coefficient, $\nu_{\rm i}$ is the ionization frequency, and $W$ is the {\it bulk drift velocity}, defined as 
\begin{equation}
W = \frac{{\rm d}}{{\rm d}t} \biggl[ \frac {\sum_{j=1}^{N(t)} z_j(t) }{N(t)} \biggr]
\label{eq:bulk}
\end{equation}
where $N(t)$ is the number of electrons in the swarm at time $t$.

In drift tubes having only two plane-parallel electrodes, the displacement current generated by the moving charge carriers is measured,\cite{Raether} which is proportional to the spatial integral of $n(z,t)$. In our system, the signal at the collector is also generated by the displacement current induced by the electrons that enter the grid-collector gap. As both of these electrodes are grounded the region between them can be assumed as free of electric field (although the field of the cathode-grid gap can penetrate to some extent between the grid wires into the grid-collector gap). A contribution of any electron upon entering this field-free region to the current signal generated at the collector is proportional to its velocity. The duration of the resulting current pulse is, however, very short, as the velocity of the electron is randomized by frequent elastic collisions. Therefore, the measured signal is proportional to the flux of the electrons entering the grid-collector gap. In the hydrodynamic regime the flux and density are proportional, i.e., in the hydrodynamic domain our system can be considered to sense the $n(z,t)$ density of the swarm, given by (\ref{eq:n}). 

Now, note that the function $n(t)$ at a given $z$ is usually {\it asymmetric}, however, $n(z)$ at any given $t$ is a {\it symmetrical}, Gaussian function. Therefore, cutting and fitting $n(z,t)$ at a sequence of $t$ values gives the mean (center-of-mass) position $\langle z(t) \rangle$. When this position is plotted as a function of time, the slope gives the bulk drift velocity, which provides a completely transparent procedure for the determination of $W$. 

\begin{figure*}[t!]
\includegraphics[width=\textwidth]{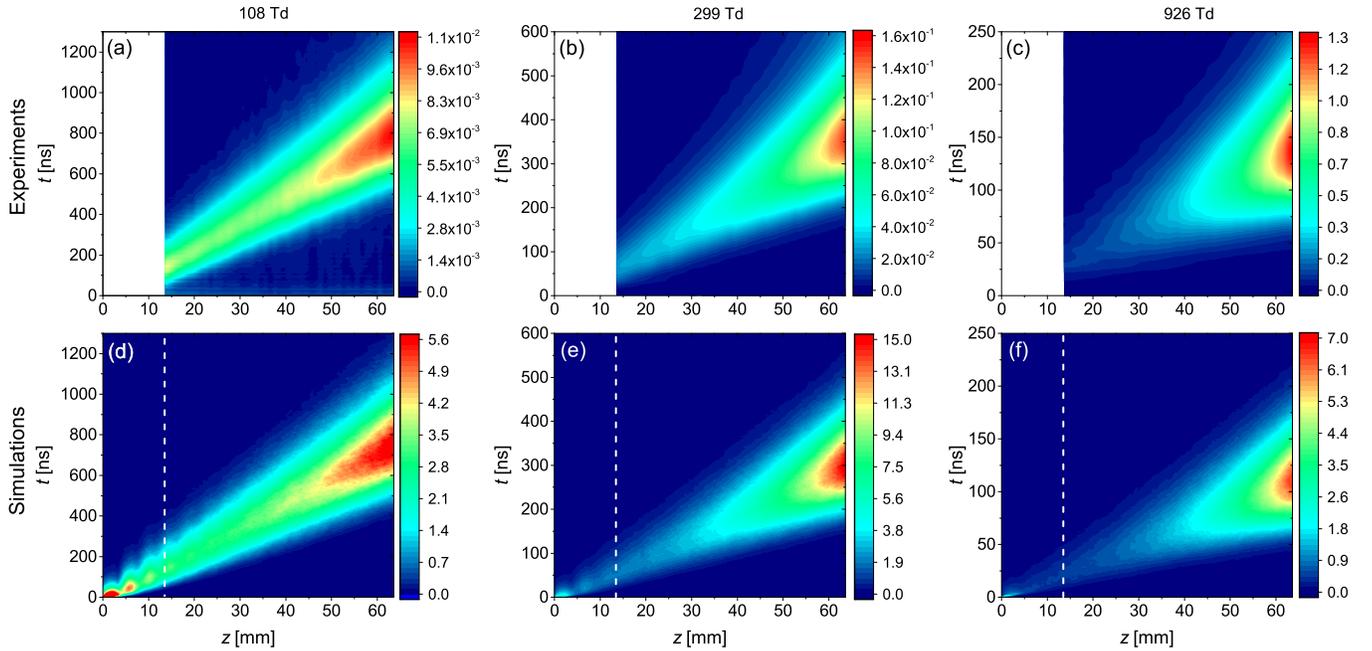}
\caption{\label{fig:xtplots}
(color online) Spatio-temporal evolution of electron swarms in argon, for $E/N$ = 108 Td (a,d), 299 Td (b,e), and 926 Td (c,f). Upper row: experimental results (measured current), lower row: computational results (electron density). The color scales give the quantities in arbitrary units, The vertical dashed white lines in the bottom row of panels denote the minimum cathode-grid position in the experiment, below which experimental data are not accessible.}
\end{figure*}  

\section{Simulation method}

The electrons' motion within the drift cell is also followed numerically, by Monte Carlo (MC) simulation,\cite{MC} which provides a fully kinetic description under the conditions of non-equilibrium transport.\cite{noneq} The cross section set for e$^-$--Ar collisions are taken from Ref.\citenum{ar-cs}. The electrons, released from the cathode due to photo-effect, are assumed to have a random kinetic energy between 0 and 1 eV. The directions of their initial velocity vectors are distributed over a half sphere. These electrons, as well as the additional electrons created in ionizing collisions are traced until they are absorbed at the cathode or at the collector electrode; their trajectories are obtained by integrating their equations of motion, 
\begin{equation}
m \ddot{\bf r}(t) = q {\bf E}, \nonumber
\end{equation}
with a time step $\Delta t$, in the presence of a stationary and homogeneous electric field ${\bf E}$. Here $m$ and $q$ are the mass and the charge of the electrons, respectively. The occurrence of collisions is checked at every time step, an electron-atom collision (in the cold gas approximation) occurs with a probability 
\begin{equation}
P(\Delta t) = 1- \exp[-N \sigma_{\rm T}(v) v \Delta t], \nonumber
\end{equation}
where $N$ is the density of the background gas atoms, $\sigma_{\rm T}(v)$ is the total electron-atom collision cross section and $v$ is the velocity of the electron. The time step $\Delta t$ has to be carefully chosen to be a small fraction of the time between collisions. The types of collisions and the scattering processes are treated in a stochastic manner, adopting the usual procedures of MC simulations.\cite{donko_psst} In the simulations we assume that all types of collisions result in isotropic scattering and, accordingly, for elastic collisions we use the elastic momentum transfer cross section. In ionizing collisions the scattered and the ejected electrons share the available kinetic energy equally. The simulation is three dimensional in both the velocity space and in the real space. (However, in the simulations we disregard the finite radial extent of the cell.) Besides their absorption at the electrode surfaces electrons may also be reflected there. Reflection of electrons is only considered at the cathode, where we assume a reflection probability\cite{Kollath} of 0.2. Due to the high transmittance of the mesh, electron reflection is neglected there. 
                   
In the simulations we compute the spatial and temporal distributions of the density and the flux of the electrons, as well as the bulk drift velocity ($W$, defined by eq.(\ref{eq:bulk})) and flux ($w$) drift velocity defined as 
\begin{equation}
w = \frac {\sum_{j=1}^{N(t)} v_{j,z}(t)} {N(t)},
\end{equation}    
where $v_{j,z}$ is the field-parallel velocity component of the $j$-th electron. $W$ and $w$ are measured at sufficiently long times and over sufficiently extended spatial domains to make sure that the equilibrium (hydrodynamic) values of the velocities are captured.
                   
\section{Results}

The swarm mapping capability of the apparatus is demonstrated in Figure~\ref{fig:xtplots}, where the spatio-temporal distribution of the experimentally measured current (top row) and the numerically obtained electron density (bottom row) are displayed for three different sets of conditions. While signs of the equilibration of the swarms near the cathode are observable in the numerical results, as periodic changes of the density seen especially at the two lower values of the reduced electric field ($E/N$ = 108 Td and 299 Td), in the domain accessible experimentally ($z \geq$ 13.6 mm) the swarms are in the hydrodynamic regime. As we shall show later the equilibration of the swarm can also be captured experimentally by choosing proper conditions.

There is a remarkable similarity between the experimental and simulated maps. All maps clearly show (i) the drift of the swarm (increase of mean position with time), (ii) the lateral diffusion (widening of the distribution along $z$ as time increases, and (iii) the presence of ionization (increasing "amplitude" with increasing position). The latter is especially sensitive to $E/N$, as the comparison of columns of Fig.~\ref{fig:xtplots} reveals.

\begin{figure}[t!]
\includegraphics[width=0.4\textwidth]{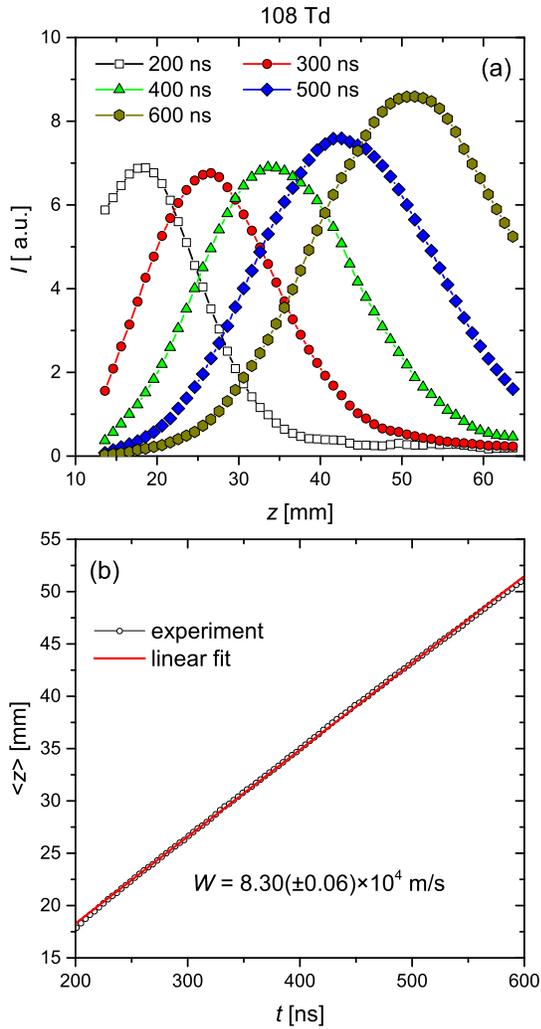}
\caption{\label{fig:cuts}
(color online) (a) Measured signal intensity corresponding to cuts of the $n(z,t)$ distributions (see eq.(\ref{eq:n})) at different times, at $E/N$ = 108 Td. (b) Illustration of the determination of the bulk drift velocity: $W$ is obtained as the slope of the linear fit to the central position of the swarm, $\langle z \rangle$, as a function of time.}
\end{figure}

Figure~\ref{fig:cuts} demonstrates the procedure for determination of the bulk drift velocity. As mentioned in Sec. II B, the ``cuts'' of the $n(z,t)$ distribution at fixed times (vertical cuts of the maps in Fig.~\ref{fig:xtplots}) yield Gaussian $n(z)$ functions. Such cuts are displayed in Fig.~\ref{fig:cuts}(a) for the case of $E/N$=108 Td. Depending on the time, the Gaussians are truncated to smaller or greater extent, due to the finite spatial domain that is accessible experimentally. As these curves are symmetrical, the mean value is found simply as the peak position following a Gaussian fit. We measure the peak position that gives $\langle z \rangle$, as a function of time, as it is illustrated in Fig.~\ref{fig:cuts}(b), which confirms that the values very closely follow a straight line, the slope of which defines the bulk drift velocity. (Small deviations from a perfect straight line fit can be attributed to the slightly inaccurate Gaussian fitting of heavily truncated distributions at ``early'' and ``late'' times, cf. Fig.~\ref{fig:cuts}(a)).

\begin{figure}[t!]
\includegraphics[width=0.4\textwidth]{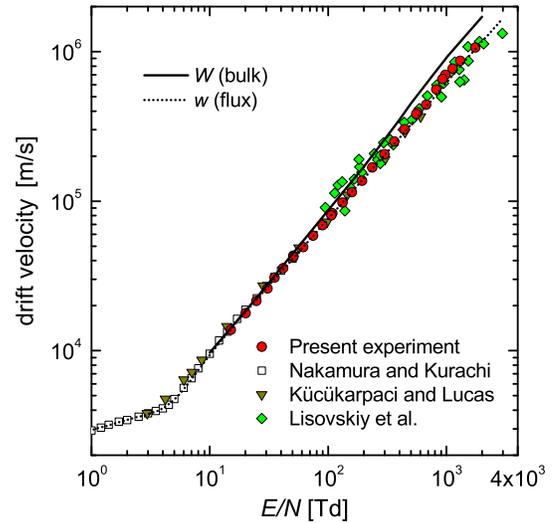}
\caption{\label{fig:bulkW}
(color online) Bulk drift velocity values determined from our mesurements, in comparison with the results of previous measurements by Nakamura and Kurachi,\cite{NK} K\"uc\"ukarpaci and Lucas\cite{KL}, and Lisovskiy {\it et al}.\cite{Lis} The solid and dotted lines, respectively, indicate the bulk ($W$) and flux ($w$) drift velocities obtained from our Monte Carlo simulations adopting the cross section set of Ref.\citenum{ar-cs}.}
\end{figure}

The results obtained for $W$ from the above measurement procedure are presented in Fig.~\ref{fig:bulkW} as a function of $E/N$, in comparison with results of previous investigations. Our data agree well with those obtained by Nakamura and Kurachi,\cite{NK} as well as K\"uc\"ukarpaci and Lucas\cite{KL}. In these experiments the drift velocity was determined for $E/N <$ 600 Td, from time-of-flight spectra measured at several gap distances in drift tubes operated with heated-filament sources of electrons. Data for higher $E/N$ values were given by Lisovskiy {\it et al}.\cite{Lis} Their data, which were derived from the radiofrequency breakdown characteristic of the gas and not in a drift tube experiment, show comparatively more scattering, but agree generally well with our results. 

Figure~\ref{fig:bulkW} also shows the results for the flux ($w$) and bulk ($W$) drift velocities, obtained from our Monte Carlo simulations. The two velocity values are identical at low $E/N$ values, where transport is conservative. Noticeable differences show up above $E/N \approx$ 50 Td where ioniziation processes become relevant. Under these (non-conservative) conditions the bulk drift velocity exceeds the flux drift velocity. The measured data follow more closely the curve for the flux drift velocity, which indicates that the cross section set of Ref.\citenum{ar-cs} does not allow an accurate reproduction of the bulk and flux drift velocity values above $\sim$ 100 Td.    

\begin{figure}[t!]
\includegraphics[width=0.41\textwidth]{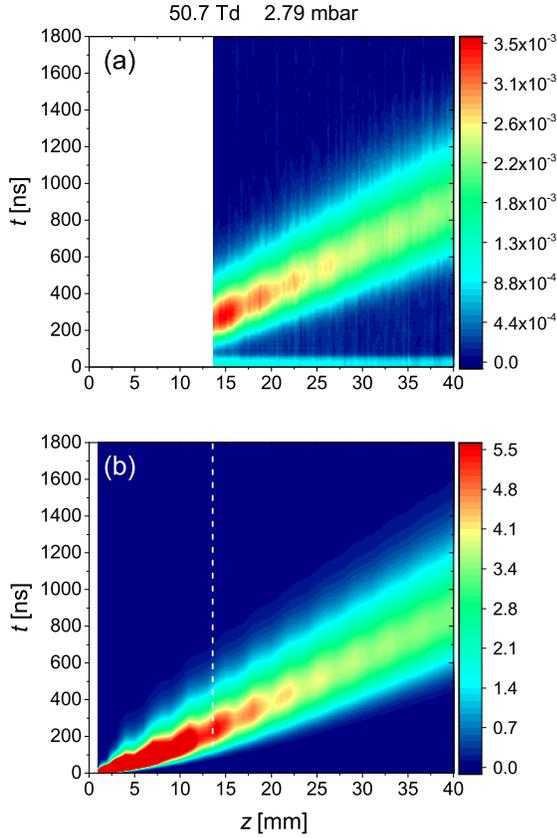}
\caption{\label{fig:equi}
(color online)  Spatio-temporal evolution of an electron swarm in argon, at $E/N$ = 50.7 Td and $p$ = 2.79 mbar: (a) experimentally recorded map of the measured flux, (b) simulated flux. Note the periodic modulation of the flux, indicative of non-equilibrium transport. The vertical dashed white line in (b) denotes the minimum cathode-grid position in the experiment, below which experimental data are not accessible. The color scale of the simulated data is saturated in this domain in order to have comparable intensities of the experimental and simulation data at $z>$13.6 mm.}
\end{figure}

Finally, in Fig.~\ref{fig:equi} we show an exemplary map recorded at conditions ($E/N$ = 50.7 Td and $p$ = 2.79 mbar) when the equilibration of the swarm requires an extended spatial domain, that not only has region-of-interest ramifications for both observation and simulation, but also allows for the observational resolution of this phenomenon. In the map 6-7 local maxima of the flux can clearly be observed. The modulation amplitude percentage of the flux decreases with increasing time, indicating the gradual establishment of the balance between the energy gain and energy loss processes (i.e., the equilibration) of the swarm. The corresponding features can be clearly identified in the numerical results. 

\section{Summary}

We report on the details and the operation of a new {\it scanning drift tube} apparatus. For the first time, to the best of our knowledge, we map and vizualize the spatiotemporal evolution of electron swarms generated by photoelectron pulses using this specially designed scanning drift tube. 

The measured swarm maps (also well reproduced by numerical simulations) allow identification of the {\it hydrodynamic}, as well as the {\it non-hydrodynamic} regimes. In the measurements we follow the progression of the center-of-mass of the swarm with time, and so, provide a transparent and accurate way for the determination of the bulk drift velocity. Besides the measurement of the bulk drift velocity based on this procedure, fitting of the ``whole'' experimentally measured distributions with the functional form (\ref{eq:n}) allows the determination of all the three transport parameters ($W$, $D_{\rm L}$, and $\nu_{\rm i}$) involved. Due to the large information content of the $n(z,t)$ distributions, such fitting results in accurate data, as will be demonstrated in a forthcoming publication for various gases.\cite{next} 

\acknowledgements

This work has been supported by the Grant OTKA K-105476. We thank T. Sz\H{u}cs and P. Hartmann for their help in the construction of the experimental setup. We thank N. Pinh\~{a}o and D. Loffhagen for useful discussions.

\end{document}